\documentclass{article}

    \PassOptionsToPackage{numbers,compress}{natbib}

    \usepackage[preprint]{nips}

\usepackage[utf8]{inputenc} 
\usepackage[T1]{fontenc}    
\usepackage{hyperref}       
\usepackage{url}            
\usepackage{booktabs}       
\usepackage{amsfonts}       
\usepackage{amsmath}
\usepackage{nicefrac}       
\usepackage{microtype}      
\usepackage{xcolor}         

\usepackage{algorithmic}
\usepackage{algorithm}
\usepackage{multirow}
\usepackage{bm}
\usepackage{subcaption}
\usepackage{graphicx}
\usepackage{enumitem}
\usepackage{wrapfig}

\title{Incorporating Pre-training Paradigm for Antibody  Sequence-Structure Co-design}

\author{Kaiyuan Gao$^{1}$\thanks{This work is conducted at Microsoft Research AI4Science.}\,\,, Lijun Wu$^2$\thanks{Corresponding author.}, Jinhua Zhu$^3$, Tianbo Peng$^4$, Yingce Xia$^2$, Liang He$^2$, Shufang Xie$^2$, \\ \textbf{Tao Qin}$^2$, \textbf{Haiguang Liu}$^2$, \textbf{Kun He}$^1$, \textbf{Tie{-}Yan Liu}$^2$\\
$^1$School of Computer Science and Technology, Huazhong University of Science and Technology;\\
$^2$Microsoft Research AI4Science;\\
$^3$CAS Key Laboratory of GIPAS, University of Science and Technology of China; \\
$^4$School of Life Sciences and Biomedical Pioneering Innovation Center, Peking University; \\
\texttt{\{im\_kai, brooklet60\}@hust.edu.cn}\\
\texttt{\{teslazhu\}@mail.ustc.edu.cn}, \quad \texttt{ptbian@pku.edu.cn}\\
\texttt{\{lijuwu, yinxia, lihe, shufxi, taoqin, haiguangliu, tyliu\}@microsoft.com}\\
}

\begin{document}

\maketitle

\begin{abstract}
Antibodies are versatile proteins that can bind to pathogens and provide effective protection for human body. Recently, deep learning-based computational antibody design has attracted popular attention since it automatically mines the antibody patterns from data that could be complementary to human experiences. However, the computational methods heavily rely on the high-quality antibody structure data, which is quite limited. Besides, the complementarity-determining region (CDR), which is the key component of an antibody that determines the specificity and binding affinity, is highly variable and hard to predict. Therefore, data limitation issue further raises the difficulty of CDR generation for antibodies. Fortunately, there exists a large amount of sequence data of antibodies that can help model the CDR and alleviate the reliance on structure data. By witnessing the success of pre-training models for protein modeling, in this paper, we develop the antibody pre-training language model and incorporate it into the (antigen-specific) antibody design model in a systemic way. Specifically, we first pre-train an antibody language model based on the sequence data, then  propose a one-shot way for sequence and structure generation of CDR to avoid the heavy cost and error propagation from an autoregressive manner, and finally leverage the pre-trained antibody model for the antigen-specific antibody generation model with some carefully designed modules. Through various experiments, we show that our method achieves superior performances over previous baselines on different tasks, such as sequence and structure generation, antigen-binding CDR-H3 design.

\end{abstract}


\section{Introduction}

\begin{wrapfigure}{r}{0.5\textwidth}
    \centering
    \includegraphics[width=1.0\linewidth]{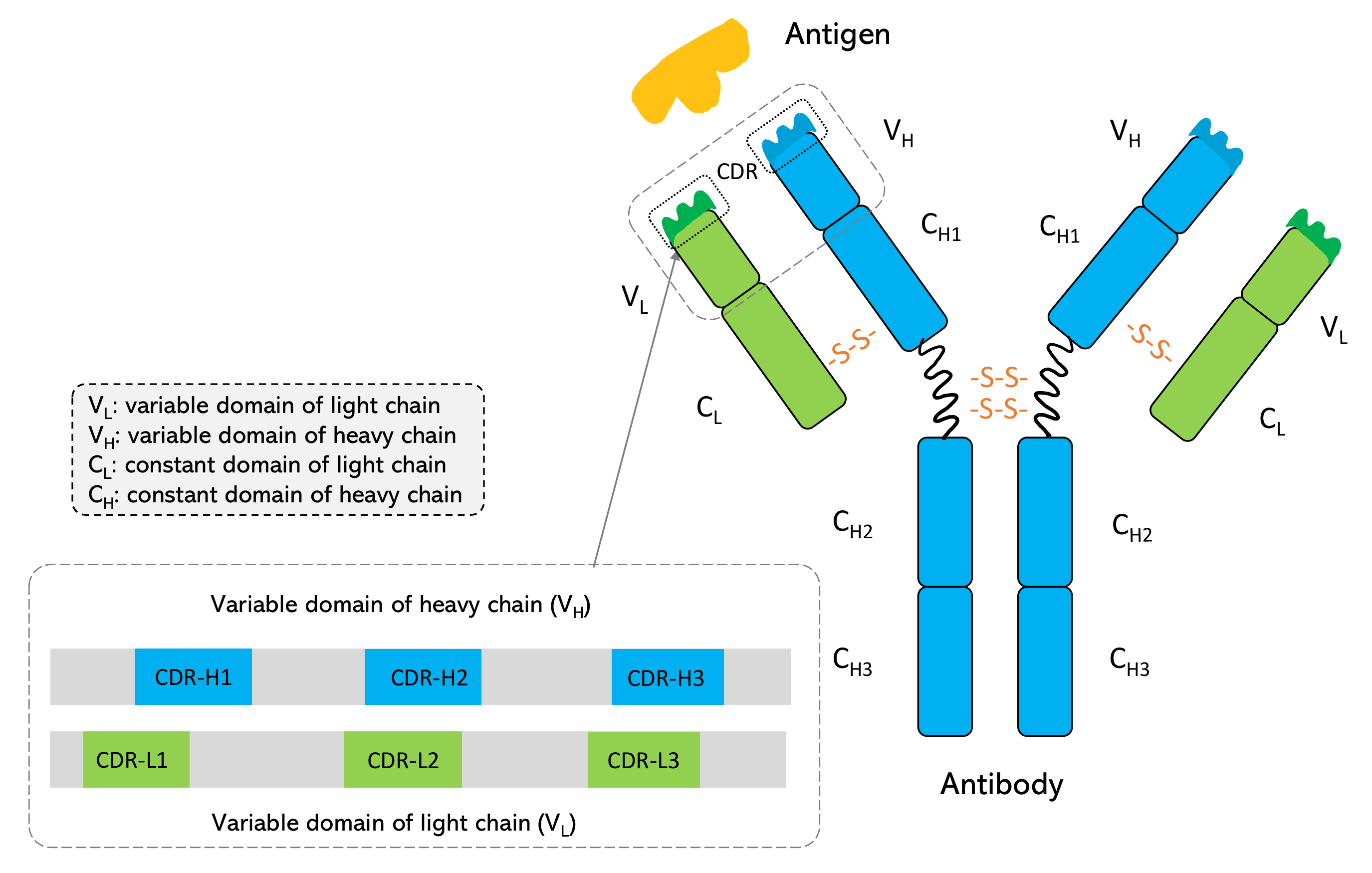}
    \caption{The Y-shape of an antibody. The CDR-H$n$ are the variable CDRs that belong to the heavy chain, which are the most important parts for binding. Below is the sequential format of the variable domains of two chains.}
    \label{fig:antibody}
\end{wrapfigure}

Antibodies are Y-shaped proteins (Figure~\ref{fig:antibody} for the overall structure of an antibody) and they are crucial biological elements in human immune system as therapeutics targeting various pathogens, treating cancer, infectious diseases and so on~\cite{scott2012antibody}. They have several characteristics. First, the antibodies have strong specificity towards the effectiveness~\cite{renart1979transfer}. Most antibodies are monoclonal that each kind of antibody usually binds to a unique type of protein (antigen). 
Second, the binding areas of antibodies are mainly determined by complementarity-determining regions (CDR), while the CDRs are highly variable with free loop and other unstructured shapes, especially the CDRs on the heavy chain~\cite{wang2007antibody}. 
Therefore, the crucial problem of antibody design mainly focuses on how to identify and design novel CDRs that can effectively and stably bind to the specific antigen. 

Recently, computational methods~\cite{gentiluomo2019application,hummer2022advances,liberis2018parapred} have been explored to automatically create the CDR sequences with desired properties, e.g., high binding affinity. Especially, deep learning has demonstrated its great potential for the antibody design, such as deep generative models~\cite{saka2021antibody}, graph neural networks~\cite{shan2022deep}. Previously, people focus more on generating the CDR sequences only \cite{melnyk2021benchmarking,zhai2011synthetic,alley2019unified,shin2021protein}. However, co-designing the sequences with their 3D structures is a more promising choice because of its realistic/practical value and it becomes a recent trend. For example,  \cite{jin2021iterative} and  \cite{jin2022antibody} both utilize the deep graph neural networks to model the antibody generation from sequence and structure. Though huge progress has been made, there still exists several key challenges. 
(1) The 3D structure data of antibodies in existing dataset is in limited size. For example, SAbDab~\cite{dunbar2014sabdab}, the most widely used dataset for antibody design, with daily collecting, it still only contains thousands of antibody structures. As for the antigen-antibody complexes, which are the key ingredients to model the interaction between antibodies and antigen, the number is even more limited. However, the antibody space of a single individual is estimated to be at least $10^{12}$~\cite{schmitz2022human}, which is far beyond the volume of existing database. Also, compared with the common applications that achieved big success, this kind of data scale is far from enough for training deep learning models 
Besides, due to the highly variable property of the CDRs, the limited existing structures further hinders the learning ability for deep models to accurately predict the CDR structures.  
(2) The current approaches for antibody design usually adopt the autoregressive generation manner~\cite{jin2021iterative}, which is to predict the amino acid type and the corresponding structure coordinates one by one. The drawbacks of such a way are obvious. On the one hand, the step-by-step amino acid generation suffers from the low efficiency that it will take multiple iterations (the length of the CDR) to complete the CDR design. On the other hand, it is also well known that this autoregressive way faces the error accumulation problem~\cite{ranzato2015sequence,wu2019beyond}, where the error from the previous step will propagate to next step during generation so that the accuracy is influenced. 
These issues lead to unsatisfactory performance on the sequence and structure co-design. For instance, the amino acid recovery (AAR) rate is only about $30\%$ for the CDR-H3 (the third CDR on the heavy chain), which has huge space for improvement. 

To address above challenges, in this paper, we propose several strategies that can help from different perspectives.
(a) Although the structured antibody data is in limited scale, there are millions of antibody sequence data that we can leverage. It is widely acknowledged that the sequence of the protein can determine its structure \cite{lesk1980different,flower1993structure}. Sequence pre-training has demonstrated its power in various domains~\cite{kenton2019bert,liu2019roberta,li2020oscar,zhou2020unified}, and the protein sequence pre-training also demonstrates its ability for protein structure prediction~\cite{wu2022high,chowdhury2022single}. Therefore, we utilize the large amount of the sequence data for antibody pre-training. By doing so, the antibody representations can be greatly enhanced and thus alleviate the problem brought by the lack of structured data.
(b) We formulate the antibody generation process in a one-shot way rather than the autoregressive manner. That is, we simultaneously generate all the amino acids of CDR at once, and the structure of the generated amino acids are also updated at one time. In this way, the error propagation problem~\cite{ranzato2015sequence} between different steps of the autoregressive way is avoided. To ensure the performance, we also refine the whole generated sequences and structures of CDR with several iterations like Alphafold2~\cite{jumper2021highly}.
(c) We incorporate the pre-trained antibody language model into the antibody design model in a careful way. Instead of only taking the prediction of the pre-trained model as initialization for the antibody design model, we introduce two integration ways to make the best use of the pre-trained model. Specifically, we use prompt tuning~\cite{lester2021power,li2021prefix} strategy to better finetune the pre-trained model, which is supposed to not only keep the ability of the pre-trained model but also provide the transferable knowledge for downstream antibody design. Besides, we systematically fuse the intermediate representations of the two models and then feed for sequence and structure prediction of CDR. Through above designed strategies, we can successfully improve the CDR generation from both efficiency and effectiveness perspectives.

Extensive experiments are conducted to verify the value of our method. Specifically, we evaluate on two generation tasks: sequence generation and structure prediction, antigen-binding CDR-H3 design. Compared to existing works, our method achieves the state-of-the-art performances from different evaluation metrics.

Our contributions are summarized as follows: (1) We propose to pre-train a sequence-based antibody language model and successfully incorporate the pre-trained model for antibody sequence-structure co-design. (2) We introduce the one-shot generation way to replace the autoregressive manner so to make the decoding process more efficient. (3) Empirical studies verify the effectiveness and efficiency of our proposed method and the state-of-the-art performances are achieved for antibody design.




\section{Related Work}

\subsection{Antibody Pre-training}
Inspired by the great success of protein pre-training~\cite{elnaggar2021prottrans,zhang2022ontoprotein,strodthoff2020udsmprot,nambiar2020transforming,wu2022sproberta,he2021pre}, a few works attempt to transfer these techniques to the antibody pre-training since the specificity of antibody~\cite{vu2022immunolingo,vu2022advancing}. ~\citet{ruffolo2021deciphering} first propose an antibody-specific language model AntiBERTy to aid understanding of immune repertoires by training on antibody sequence data. They find the model can cluster antibodies into trajectories resembling affinity maturation and identify key binding residues. ~\citet{leem2022deciphering} introduce AntiBERTa, an antibody-specific bidirectional encoder representation from Transformers, the success is demonstrated by the leading results in antibody binding site prediction and paratope position prediction. ~\citet{olsen2022ablang} pre-train two antibody models, an Ablang-H model trained on the heavy chain and an Ablang-L model trained on the light chain of antibody. They show the model power on restoring missing residues in antibody sequence data, which also surpasses the general protein pre-training model ESM-1b~\cite{rives2021biological}. ~\citet{shuai2021generative} introduce a deep generative language model for generating synthetic libraries by re-designing variable-length spans of antibody sequences. Our work differs from above works that we incorporate the pre-trained sequence model for both sequence and structure predictions of antibody.

\subsection{Antibody Design}
Antibody design is special to the general protein design. Protein design~\cite{o2018spin2,ingraham2019generative,strokach2020fast,cao2021fold2seq,karimi2020novo} mainly focuses on the sequence, generation problem that conditioned on a known 3D structures, while antibody design is based on the assumption that both the sequence and the structure of antibody are unknown.  
Traditional computational antibody design methods mainly utilize the energy function optimization, which use the physics inspired methodology to optimize the sequence and structure of the antibody to reach a minimal energy state~\cite{li2014optmaven,chowdhury2018optmaven,lapidoth2015abdesign,adolf2018rosettaantibodydesign}. Monte Carlo simulation process is a typical adopted way to search over the energy space. However, these methods seriously suffer from the high cost and low computation efficiency. 
Recently, the deep learning methods have attracted much attention for antibody design and different generative models have been proposed. Apart from the common works that only try to predict the antibody sequences~\cite{alley2019unified,shin2021protein,saka2021antibody,akbar2022silico}, co-predicting the sentence and the structure of CDRs is more promising but also challenging. How to encode both the 3D structure and sequence information and keep the specific property, e.g., equivariance, are important topics in this area. ~\citet{jin2021iterative} propose a RefineGNN model to encode the graph features using graph neural network and generate the amino acid sequences and structures in an autoregressive refinement manner. However, RefineGNN only considers the antibody itself without the specificity of antigen-antibody complex. Hence, ~\citet{jin2022antibody} further introduce HERN to model antibody-antigen docking and design via hierarchical equivariant refinement. HERN employs hierarchical message passing networks to encode both atoms and residues, the prediction is also performed in an autoregressive manner. Differently, our work co-design the CDRs in a one-shot way to avoid the autoregressive generation.

\begin{figure*}[t]
    \centering
    \includegraphics[width=0.95\linewidth]{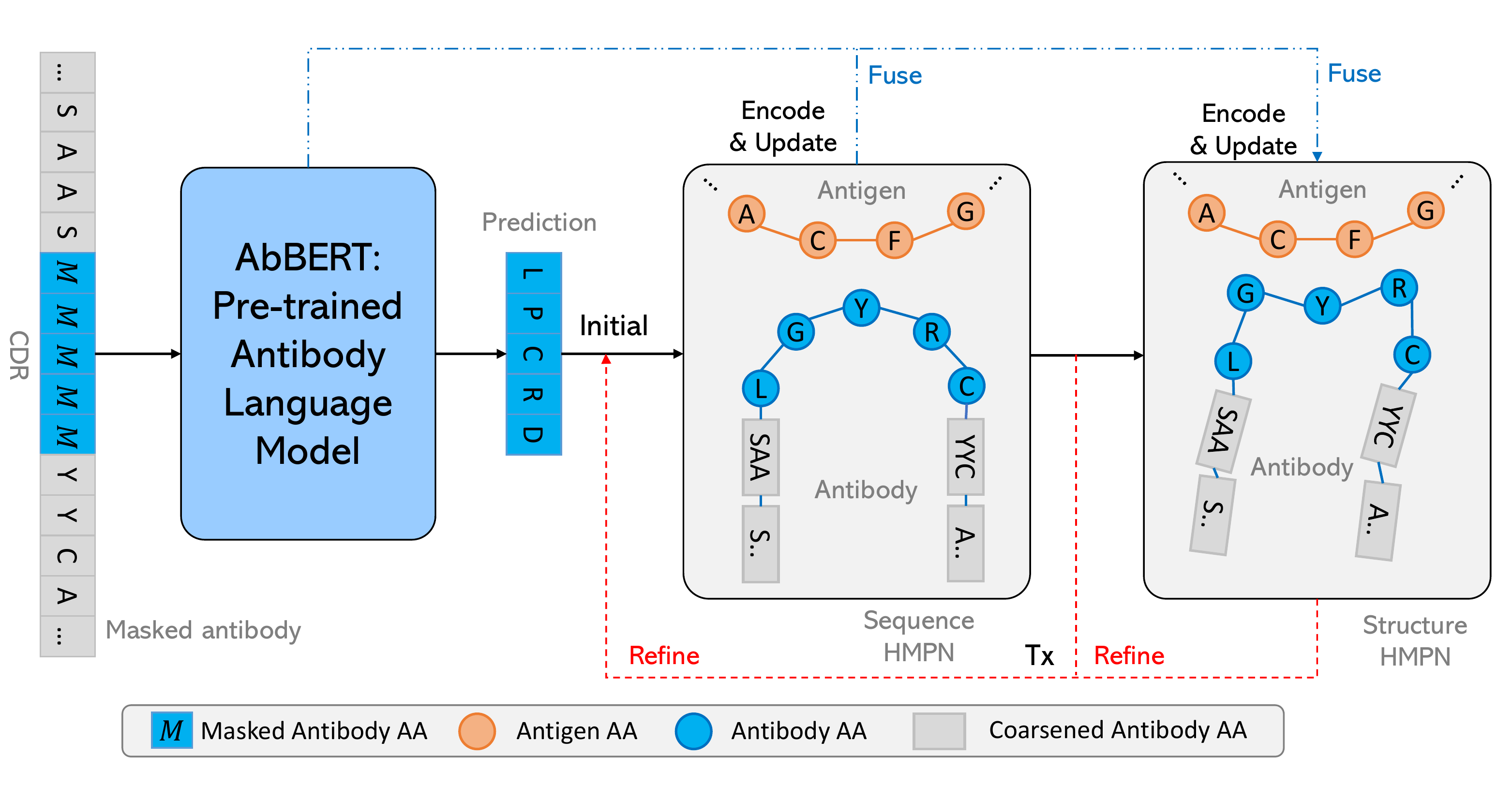}
    \caption{The overall framework of our method. The AbBERT is the pre-trained antibody model. Its `soft' prediction (for clear illustration, we draw hard prediction here, details in Section~\ref{sec:co-design})will be fed into the sequence HMPN $\mathcal{H}_{seq}$, after encoding and generating the updated sequence, structure HMPN $\mathcal{H}_{str}$ encodes the updated graph and then predict the structures. The sequence and structure prediction iteratively refine $T$ times.}
    \label{fig:framework}
\end{figure*}

\section{Methods}
In this section, we introduce our method in detail, with the background introduction at first, and then our antibody pre-training, co-designed model and the pre-training leveraging method, finally we provide the overall algorithm and also some discussions. 

\subsection{Background and Overview}
An antibody consists of two symmetric parts in a Y-shape, which is shown in Figure~\ref{fig:antibody}. In each part, there are two chains, a heavy chain and a light chain, each composed of one variable domain ($V_H$, $V_L$) and constant domains. The variable domain contains a framework region and three complementarity-determining regions (CDRs). These CDRs are the most important regions that can determine the binding affinity and the binding sites are located in CDRs. The CDRs on the heavy chain are denoted as CDR-H1/H2/CDR-H3, each filling with contiguous subsequences. Among them, CDR-H3 plays the most crucial role and with high variability, hence how to precisely predict CDRs, especially CDR-H3 is the main focus of antibody design. Following existing works~\cite{jin2021iterative,shin2021protein}, we define the antibody design as a generation task on the CDR-H1/H2/H3. In our work, we consider the design conditioned on the framework region and also the specific antigen.

We follow previous works to use deep graph generative method~\cite{jin2021iterative,jin2022antibody}. For simplicity, below we introduce the antigen-conditioned antibody generation~\cite{jin2022antibody}. When no antigen considered, it is then framework conditioned.
The binding interface of an antigen-antibody complex is composed of an epitope (in antigen) and a paratope (in antibody). The paratope is a sequence of residues in the CDR and the epitope is the sequence of residues that close to the paratope. 
Given an antibody $Ab$ and an antigen $Ag$ in complex\footnote{In the scenario of antibody design, an antibody is usually simply defined as the variable domain in heavy/chain that contains CDR and framework region, since the remaining domains of the chain are constant.}, the paratope and epitope residue sequences are denoted as $Ab^p = [Ab^p_i]_{i=1}^n$ and $Ag^e = [Ag^e_j]_{j=1}^m$, the rest framework region of $Ab$ is $Ab^f$. Since the binding occurs on the epitope of antigen, we ignore other parts in $Ag$ and only consider $Ag^e$.
As for the structures, we take the 3D structures of epitope, paratope and the paratope-epitope interface. Different from~\citet{jin2022antibody}, we also consider the framework region to enrich the context information and benefit the generation. Hence the whole antibody is considered. The structure of each part is then described as point clouds of atoms. 

In an overview, we mainly follow~\citet{jin2022antibody} to set the antibody design as a sequence generation and then structure prediction process.
The overall framework of our work is shown in Figure~\ref{fig:framework}, which consists of a pre-trained antibody language model AbBERT, a hierarchical message passing network (HMPN) for sequence encoding and generation $\mathcal{H}_{seq}$, and another HMPN for structure prediction $\mathcal{H}_{strc}$. We name our framework as \textit{AbBERT-HMPN}.
The encoding/decoding module and the specific model architecture are taken from~\citet{jin2022antibody} with necessary modifications from our innovation. 
The detailed models and the updates will be introduced in Section~\ref{sec:co-design}. 

\subsection{AbBERT: Antibody Pre-training}
As we discussed, though the structure antibody data is quite limited, it is lucky that large amount of antibody sequences are collected and existed on the web. Previous works on protein presentation learning have demonstrated the great power of pre-training language model, we hence introduce the sequence pre-training on antibody to learn expressive representations for antibody design, especially for the paratope sequence prediction. We expect our sequence antibody pre-trained model, AbBERT, can also benefit the structure prediction, where the motivation is from the widely acknowledged sense that the sequence of the protein can decide its structure and function.

\begin{figure*}
    \centering 
    \includegraphics[width=0.95\linewidth]{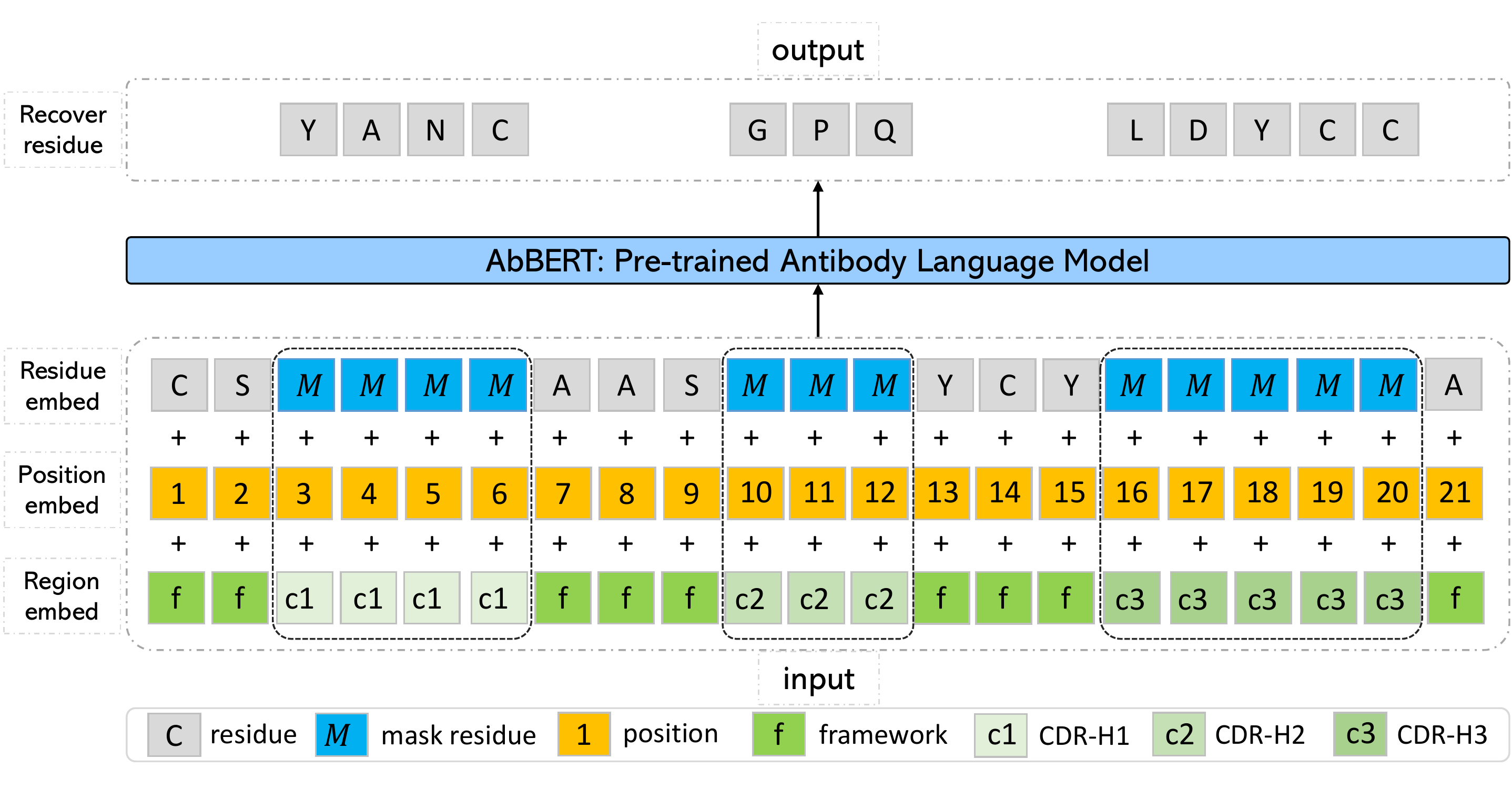}
    \caption{The AbBERT pre-training. The masking is only operated on the CDRs. To indicate the different regions of the variable domain, we set separate region embedding as input for framework (f), CDR-H1 (c1)/H2 (c2)/H3 (c3) regions accordingly.}
    \label{fig:pretrain}
\end{figure*}

Our antibody pre-training follows the style of BERT~\cite{kenton2019bert}, where the pre-training objective is masked language modeling. However, different from the general text sentence pre-training or the general protein pre-training, the antibody sequence is pretty unique as we introduced before. Hence our antibody pre-training distinguishes from them in two views. (1) The antibody contains a heavy and a light chain which are very different. Since the heavy chain is a more critical sequence, we simply take the sequence data from heavy chain for pre-training. Furthermore, along the whole heavy chain, only the variable domain is what we cared most, the pre-training data is specified to the variable domain in heavy chain ($V_H$). (2) The different regions of the variable domain also motivate us to pay different attentions to these regions. As the CDRs are the ones that can determine the binding affinity for the antibody, and the framework region serves as the contextual anchor, we set up the masked positions to be the CDRs of the variable domain, and the goal is to recover the amino acids of these masked ones in CDRs.

Specifically, given the antibody sequence $Ab=[Ab_1,...,Ab_N]$, where $N$ is the number of residues and $Ab_i$ is $i$-th residue. Instead of random masking, we look at the three CDR continuous subsequences, and we randomly select several residues in these CDR subsequences with some probability $p_m$ and replace these selected residues to be \texttt{[MASK]}. 
Then we will use the expressive Transformer~\cite{vaswani2017attention} model to encode the masked input antibody sequence with the goal of recovering the masked residues. 
Denote the original residues for the masked ones as $Ab_m$, the pre-training objective is to maximize the following log likelihood, 
\begin{equation}
\label{eqn:abbert}
    L_m(Ab) = \sum\nolimits_{m=1}^M \log P(Ab_m|\hat{Ab}),
\end{equation}
where $M$ is the number of \texttt{MASK} residues and $\hat{Ab}$ is the antibody sequence with the masked residues. 

The illustration of our antibody pre-training is presented in Figure~\ref{fig:pretrain}. There are two different operations compared to the general pre-training~\cite{kenton2019bert,liu2019roberta}. (1) First, the specific functional regions of the variable domain is clearly split, hence besides the residue embedding (similar to the token embedding) and the position embedding, we add an extra region embedding for each residue as the input, e.g., the $f, c1$ in Figure~\ref{fig:pretrain}. The four region embeddings $f, c1, c2, c3$ are constructed for the framework region, CDR-H1, CDR-H2, and CDR-H3 regions accordingly. 
(2) Second, since we only take the CDRs as the mask candidates, different from the $p_m=0.15$ setting in general pre-training, the mask ratio of these reduced candidate space (e.g., CDRs only occupy about $15\%$ of $V_H$) may require large ratio value for masking. Therefore, we configure $p_m$ to be $0.5$ or $1.0$ for masking CDRs. The effect of these different $p_m$ is studied in our experiments (see Section~\ref{sec:pretrain_effect}). 

After pre-training, our antibody specific language model AbBERT is expected to capture the common knowledge and have expressive representation for antibodies so as to benefit the antibody design in the later stage (see Section~\ref{sec:incorporate}). One notable point is that, thanks to the bidirectional property of the BERT, the recovering of these masked residues are outputted by AbBERT in a one-shot way (simultaneously predict all residues), which ensures not only a good CDR recovering accuracy (from the power of AbBERT) but also an efficient initialization for later antibody design. 

\subsection{Antibody Sequence-structure Co-design}
\label{sec:co-design}
We treat the co-design as a 3D point cloud completion, or a 3D graph node and structure completion task. The interfaced antibody structures are predicted and the paratope (CDR) sequences are generated. 
We use hierarchical message passing networks (HMPN) introduced in~\citet{jin2022antibody} for encoding and predicting the sequence and structure separately with two HMPN ($\mathcal{H}_{seq}$, $\mathcal{H}_{str}$).
As briefly introduced before, our work considers the epitope, paratope and framework for co-design, rather than the antigen and paratope in~\citet{jin2022antibody}. Thus, our encoding and decoding processes differs from~\citet{jin2022antibody} with necessary modifications. 

\subsubsection{Encoding}
The encoding part in HMPN consists of an atom-level encoding and a residue-level interface encoding to form the hierarchical encoded features for epitope and antibody. The atom-level interface encoding extracts the fine-grained features from the backbone atoms and the side-chains, while the residue-level encoding only captures the backbone $C_{\alpha}$ atoms. 
Since we consider the framework region as conditioned context and the framework is usually much longer than CDR, to save the computational cost, we use the coarse-grained encoding proposed by~\citet{jin2021iterative}. The framework residues are clustered into residue blocks. The block embedding is the mean of its residue embeddings and the block coordinate is the mean coordinate of its residues,
\begin{equation}
    E(b_i) = \sum\nolimits_{Ab^f_j\in b_i} E(Ab^f_j)/b, \quad z(b_{i,c}) = \sum\nolimits_{Ab^f_j} z(Ab^f_{j,c})/b,
\end{equation}
where $b_i$ is the block and $b$ is the residue size, $E, z$ stands for the embedding and coordinate for atom, $c$ is the type of atom. 

Before introducing the details of the atom-level encoding and the residual-level encoding, let us first describe the initial state of the CDR that we need to predict. Due to the unknown residues and structures, the residues are first initialized by a `soft' assignment from a learnable feed-forward network in~\citet{jin2022antibody} instead of a hard guess. In our work, thanks to the pre-trained AbBERT, the soft assignment of $i$-th residue is $P(Ab^p_i|\hat{Ab})$ from Eqn.(\ref{eqn:abbert}), then the node feature is initialized as 
\begin{equation}
    \label{eqn:initial}
    f(Ab^p_i) = \sum\nolimits_k P(Ab^p_i|\hat{Ab})[k]f(k),
\end{equation}
where $f(k)$ (introduced below) is the pre-defined amino acid feature for residue type $k$.
For the structures, ~\citet{jin2022antibody} use a complex distance-based initialization method, instead, we utilize a uniform split between the residue before and after the CDRs of the framework. For example, in Figure~\ref{fig:pretrain}, for CDR-H2, with the coordinates of previous $S$ and later $Y$ denoted as $z(S)$ and $z(Y)$, the initialized coordinates for the inner three residues are $[z(S)-z(Y)]/4, [z(S)-z(Y)]*2/4, [z(S)-z(Y)]*3/4$ accordingly. With these initialized residues and coordinates, we now introduce the atom-level and residue-level encoding, most of which are following~\citet{jin2022antibody}.

\noindent{\bf Atom-level Encoding.}
The atom-level encoding encodes all the atoms in the graph. For each atom, the node feature is a one-hot encoding of its atom type. The edge feature is the distance encoded by \texttt{RBF} between two atoms ($Ab^p_{i,k}, Ag^e_{j,l}$) in a radial basis, $f(Ab^p_{i,k}, Ag^e_{j,l}) = \texttt{RBF}(||z(Ab^p_{i,k})-z(Ag^e_{j,l})||)$. Then a MPN is used to learn the feature vectors $h(Ab^p_{i,k}), h(Ag^e_{j,l})$ for atom $Ab^p_{i,k}$ and atom $Ag^e_{j,l}$.

\noindent{\bf Residue-level Encoding.}
The residue-level encoding encodes only the $C_{\alpha}$ atoms in each residue. For each residue, the amino acid feature, e.g., $f(Ab^p_i)$, is first pre-defined by its dihedral angles, polarity, hydropathy and so on (in Appendix~\ref{sec:model_appendix}). Then the pre-defined feature will concatenate with the sum of the atom feature vectors that outputted from atom-level encoding to form the hierarchical residue node representation, which are defined as:
\begin{equation}
    \bar{f}(Ab^p_i) = f(Ab^p_i)\oplus \sum\nolimits_k h(Ab^p_{i, k}), 
    \bar{f}(Ab^e_j) = f(Ag^e_j)\oplus \sum\nolimits_l h(Ag^e_{j, l}), 
\end{equation}
where $\oplus$ is concatenation.  
As for the edge feature, e.g., $f(Ab^p_i, Ab^p_j)$, it is defined as the one that contains the distance, direction, and orientation between two residues, e.g., $Ab^p_i, Ab^p_j$. With the node feature and edge feature for residues, another MPN is introduced to learn the residue-level hidden representations for epitope and antibody jointly.

\subsubsection{Decoding}
After hierarchically encoding the residue and atom representations, we can generate the amino acids for CDRs and predict the structures. As shown in Figure~\ref{fig:framework}, we first use $\mathcal{H}_{seq}$ to generate the amino acids, then the generated amino acids will be updated to the $\mathcal{H}_{str}$ for structure prediction.

One big difference between our decoding and~\citet{jin2022antibody} is that we adopt an \textit{one-shot} manner instead of the autoregressive way adopted by~\citet{jin2022antibody}. The autoregressive generation suffers from two problems. One is the decoding efficiency. The step-by-step decoding is slow, which requires $n$ (length of the CDRs) times decoding, while one-shot decoding only needs one time. Another bad point is the error accumulation problem. During the generation steps, the error from last decoded step will propagate to the current decoding step, which causes the error to be enlarged and resulting wrong predictions, especially for later steps. Though our one-shot manner avoids the error accumulation problem, it can not ensure the accuracy of simultaneously generated residues. Therefore, we add $T$ refinement steps for the one-shot predicted residues and structures to improve the performance. Since $T$ is much smaller than $n$, the efficiency is guaranteed. 

\noindent{\bf Sequence Generation.}
The sequence generation step takes the encoded hidden representations. Then the amino acid prediction is performed as a multi-class classification by $\mathcal{H}_{seq}$,
\begin{equation}
    \label{eqn:softmax}
    P^{t}(Ab^p_i) = \texttt{softmax}(W_s h^{t-1}(Ab^p_i)), 
\end{equation}
where $h^{t-1}$ is the encoded representation of $t-1$ refinement step. Similar to the initial state of the residue types, we use the `soft' assignment between the first and the $t-1$ refinement steps, and use this softened version for $\mathcal{H}_{str}$ encoding, only the last step $T$ will output one hard prediction that sampled from the probability for determining the specific amino acid. 

\begin{figure}
    \centering 
    \includegraphics[width=0.85\linewidth]{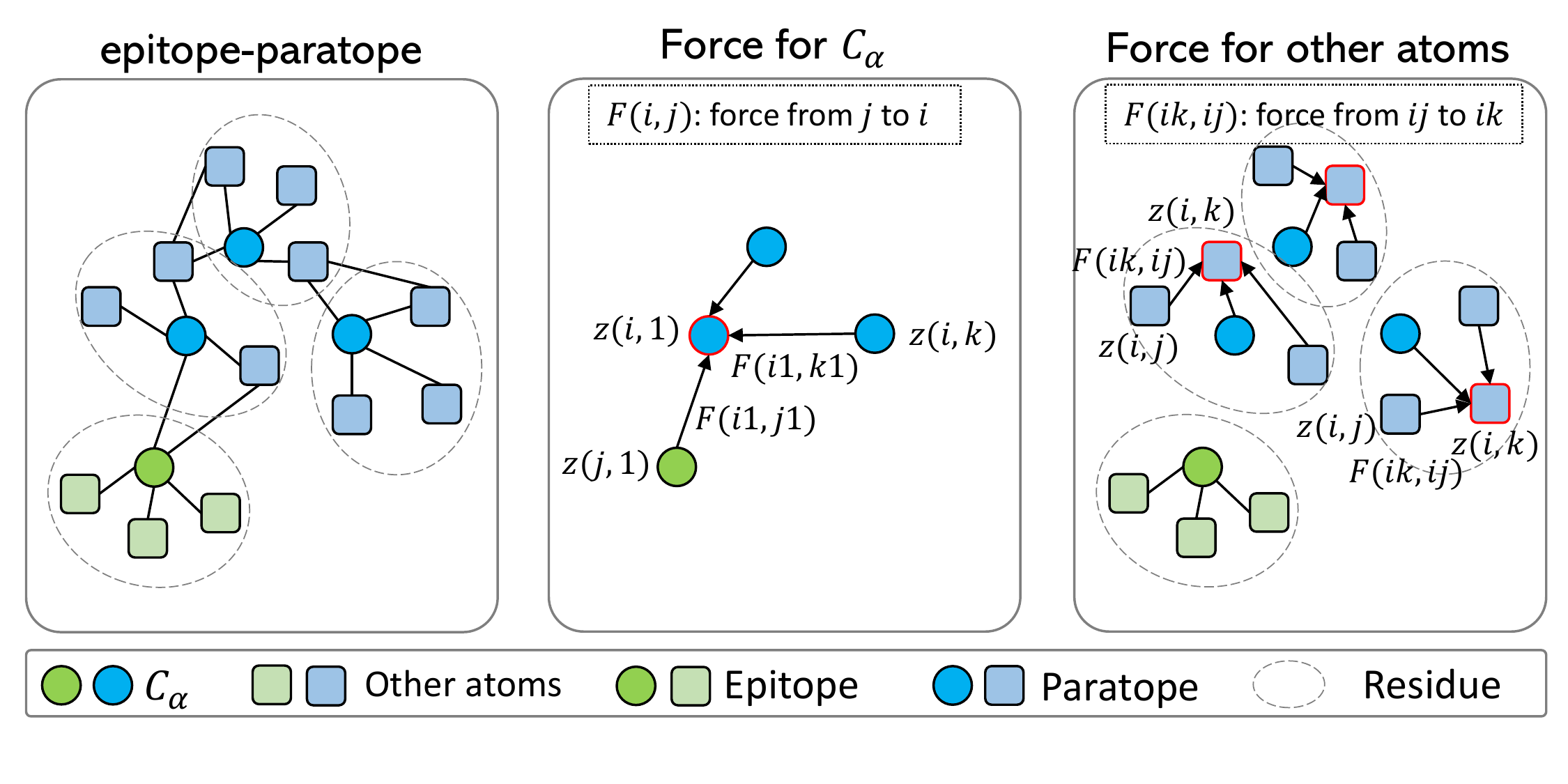}
    \caption{Two kinds of force calculation for the complex (left figure). One is between $C_{\alpha}$ atoms (middle figure), and the one one is between atoms in the same residue (right figure).}
    \label{fig:structure_update}
\end{figure}

\noindent{\bf Structure Decoding.}
As discussed in~\cite{jin2022antibody}, the structure update must maintain the equivariance w.r.t. the rotation and translation of epitope. Hence \textit{force} prediction is utilized instead of the direct coordinate prediction. The force is computed by the residue and atom hidden representations, e.g., $h(Ab^p_{i}), h(Ab^p_{i, k})$.
Specifically, the forces between $C_{\alpha}$ atoms and other atoms are separately calculated for updating the coordinates. The force between nearest $C_{\alpha}$ atoms are calculated to update the paratope $C_{\alpha}$ coordinates.
For other atoms, their coordinates are updated according to the force calculated between the atoms in the same residue. Noting that the structure update only performs on the four backbone atoms for paratope residues. The detailed calculation can be found in Appendix~\ref{sec:model_appendix} and we give a visual illustration in Figure~\ref{fig:structure_update}.
The updated structures in the $t-1$ refinement step will then feedback to the $\mathcal{H}_{seq}$ for further refinement. 

\subsection{Incorporating Pre-training}
\label{sec:incorporate}
Generally speaking, we introduce two ways to incorporate the pre-trained AbBERT for antibody design.
(1) First, we have stated in above subsection that our pre-trained AbBERT provides good initialization for $\mathcal{H}_{seq}$ and $\mathcal{H}_{str}$. The specific initialized way is shown in Eqn.(\ref{eqn:initial}), which is to give a soft probability from AbBERT. This soft initialization is supposed to be better than random guess.
(2) Second, for both $\mathcal{H}_{seq}$ and $\mathcal{H}_{str}$, we fuse the hidden representations from AbBERT with the residue representations from the encoded MPNs. That is, for the residue hidden encoding, such as $h^{t-1}(Ab^p_i)$ in Eqn.(\ref{eqn:softmax}) and the hidden $h^B(\hat{Ab}^p_i)$ before the last prediction layer from AbBERT, we simply concatenate them together and pass a linear transformation, 
\begin{equation}
    \bar{h}^{t-1}(Ab^p_i) = W_b[h^{t-1}(Ab^p_i)\oplus h^B(\hat{Ab}^p_i)],
\end{equation}
where $\oplus$ is the concatenation operation, and the fused representation $\bar{h}^{t-1}(Ab^p_i)$ will be used in later sequence generation and structure force prediction.

There are two notable points when incorporating AbBERT. (1) To save the computational cost, in each refinement step $t$, the representation from AbBERT is fixed as $h^B(\hat{Ab}^p_i)$ without updating, this can stop the multiple gradient backpropagation in the refinement steps and we find this impacts little to the performance. (2) During training, our pre-trained AbBERT is finetuned in a prefix-tuning~\cite{li2021prefix} style. That is, some randomized key and value tokens (prefix) are appended in each attention layer of AbBERT, the parameters of the pre-trained AbBERT are fixed and only the appended prefixes are finetuned. In such a way, the knowledge in AbBERT are transferred to the design model with little training cost. 
We compare the different finetuning ways in Section~\ref{sec:incor_study}.

\subsection{Summary and Algorithm}
We first introduce the training losses of our AbBERT-HMPN, which include the sequence generation loss and structure prediction loss~\cite{jin2022antibody}. The sequence loss is the commonly-adopted cross entropy loss between the predicted probabilities and the ground-truth sequence. For structure loss, they are defined as the Huber loss of the pairwise distances between all predicted atoms $||z(Ab_{i,k}) - z(Ag^e_{j,l})||$ and the ground-truth distances. 

For a quick summary of our pipeline (shown in Figure~\ref{fig:framework}) and the decoding algorithm (shown in Algorithm~\ref{alg:algorithm}), we first use AbBERT to encode the masked CDR antibody sequence, and then use the output probability and the initialized structure to initialize the graph state, then the $\mathcal{H}_{seq}$ and $\mathcal{H}_{str}$ will encode the graph, the sequence is first generated by $\mathcal{H}_{seq}$ and then the structure is predicted by $\mathcal{H}_{str}$. The sequence and structure decoding will iterate $T$ steps and finally obtain the refined sequence and structure of CDR.

\begin{algorithm}[!t]
\caption{Decoding of AbBERT-HMPN}
\label{alg:algorithm}
Finetuned AbBERT $\mathcal{M}_B$, sequence HMPN $\mathcal{H}_{seq}$ and structure HMPN $\mathcal{H}_{str}$
\begin{algorithmic}[1]
\STATE Predict the initial sequence from the output of $\mathcal{M}_B$, initialize the structure and build graph $G^0$
\FOR{t=0 to T-1} 
\STATE Encode the graph $G^t$ with $\mathcal{H}_{seq}$
\STATE Generate the sequence with $\mathcal{H}_{seq}$
\STATE Update the sequence and encode the sequence updated graph with $\mathcal{H}_{str}$
\STATE Predict the structure with $\mathcal{H}_{str}$
\STATE Update the structure and build the new graph $G^{t+1}$
\ENDFOR
\STATE Output the sequence and structure from $G^T$
\end{algorithmic}
\end{algorithm}

\subsection{Discussion}
\label{sec:discuss}
One can see that our work is based on~\citet{jin2022antibody}. Apart from incorporating the AbBERT in the pipeline, our work differs from the following points. (1) The context information. Unlike their work, for the antigen we only consider the epitope instead of the entire antigen, and for antibody we add the framework region besides the CDRs. Since the design is on the antibody, we believe more context information from the antibody is beneficial. (2) The graph initialization. We use our AbBERT output as the residue initialization rather than random guess. Besides, our uniformly split structure initialization also distinguishes from the complex one in~\citet{jin2022antibody}. (3) The decoding strategy. Differing from the inefficient autoregressive generation way, our one-shot decoding greatly speed up the inference process with higher accuracy. 

\section{Experiments}

\subsection{Overview Settings}
To evaluate our approach, we conduct experiments on two generation tasks. Following previous works~\cite{jin2021iterative,jin2022antibody}, the first setting is sequence generation and structure prediction (Section~\ref{sec:exp1}), the second task is antigen-binding CDR-H3 design (Section~\ref{sec:exp2}).


\noindent{\bf Baseline Models.}
Since our work is upon~\citet{jin2022antibody}, most of our compared baseline models are following theirs, including (1) RosettaAntibodyDesign (RAbD)~\cite{adolf2018rosettaantibodydesign}, which is a physics-based method used for sequence generation and energy minimization. (2) A LSTM-based model that only works on sequence generation without structure information~\cite{saka2021antibody,akbar2022silico}. (3) Sequence model. A 3D structure-based MPN encoder and an RNN-based sequence decoder, which is implemented by~\citet{jin2022antibody}. (4) AR-GNN. An autoregressive-based graph generation model~\cite{jin2020multi} that predicts the amino acid and then the edge iteratively to form the graph. (5) RefineGNN~\cite{jin2021iterative}. A GNN-based model that encodes the residue and structure information and co-designs the sequence and structure with iterative refinement. Noting that this model is antibody only. (6) HERN~\cite{jin2022antibody}. The antigen-specific antibody design model with hierarchical equivariant message passing neural networks, which also uses autoregressive decoding and iterative refinement. 

\noindent{\bf Model Settings.}
Our pre-trained AbBERT is a $12$-layer Transformer model, each MPN in the $\mathcal{H}_{seq}$ and $\mathcal{H}_{str}$ consists of $4$ message passing layers with hidden dimension $256$. The refinement steps $T$ in our method is 5. The training details are in Appendix~\ref{sec:model_appendix}. 

\subsection{Pre-training}
For pre-training AbBERT, we take the existing antibody sequence data from Observed Antibody Space database (OAS)~\cite{olsen2022observed}. We follow~\cite{Bachas2022.08.16.504181} to preprocess the OAS dataset, which contains sequence filtration and clustering according to the full-length Fv amino acid sequences. We only take the sequences from heavy chain for pre-training. After preprocessing, there are 118825825 sequences remained. We sample 50 millions for pre-training.
Then we use a BERT$_\texttt{{base}}$ configuration to train AbBERT. The details about the data, pre-training model, and the performance are in Appendix~\ref{sec:pretrain_appendix}.

\subsection{Sequence and Structure Prediction}
\label{sec:exp1}
\noindent{\bf Data.} 
This task aims to evaluate the general ability of the generative model. The data is from Structural Antibody Database (SAbDab)~\cite{dunbar2014sabdab}. To make a fair comparison, we follow~\citet{jin2021iterative} that only use the antibody sequences without antibody in this task. We directly take the data that processed by~\citet{jin2021iterative}, where the data is split to train/valid/test set with $8:1:1$ ratio according to the CDR cluster. The number of clusters for CDR-H1, CDR-H2 and CDR-H3 are $1266$, $1564$ and $2325$ accordingly. 

\noindent{\bf Results.} 
The evaluation metrics for this task are perplexity (PPL), root mean square deviation (RMSD), and amino acid recovery (AAR) between the predicted antibody and the ground-truth. The PPL and AAR are used to measure the sequence generation accuracy and RMSD is used to measure the structure prediction accuracy. The results of this task is shown in Table~\ref{tab:exp1}. From the results, we can see that our sequence generation and structure prediction are significantly better on the three metrics than previous works. Specifically, the PPL and RMSD are reduced in a large margin on three CDRs, on CDR-H1/H2, the RMSD is even smaller than $1.0$. The AAR accuracy is hugely improved, for example, on CDR-H1, about $20$ points accuracy improvement is achieved, which greatly prove the strong modeling ability of our method.

\begin{table*}[t]
  \centering
  \caption{Results of the sequence generation and structure prediction task on CDR-H1/H2/H3. PPL, AAR, and RMSD stand for perplexity, amino acid recovery and root mean square deviation. The results of the baselines are reproduced by the released code and models from~\citet{jin2021iterative}.}
  \scalebox{0.88}{
  \begin{tabular}{l ccc| ccc | ccc}
    \toprule
    \multirow{2}{*}{\textbf{Method}} & \multicolumn{3}{c}{\textbf{CDR-H1}} & \multicolumn{3}{c}{\textbf{CDR-H2}} & \multicolumn{3}{c}{\textbf{CDR-H3}}\\
    \cmidrule{2-10}
     & \textbf{PPL}$\downarrow$ & \textbf{RMSD}$\downarrow$ & \textbf{AAR$\uparrow$} & \textbf{PPL}$\downarrow$ & \textbf{RMSD}$\downarrow$ & \textbf{AAR$\uparrow$} & \textbf{PPL}$\downarrow$ & \textbf{RMSD}$\downarrow$ & \textbf{AAR$\uparrow$} \\
    \midrule
     LSTM & 6.79 & - & - & 7.21 & - & - & 9.70 & -  & - \\
     AR-GNN & 6.99 & 2.87 & 41.88\% & 6.84 & 2.34 & 41.18\% & 9.23 & 3.19 & 18.93\% \\
     RefineGNN & 3.90 & 1.39 & 34.53\% & 5.15 & 1.71 & 29.68\% & 7.25 & 2.62 & 24.22\% \\
     \midrule
     AbBERT-HMPN & \textbf{2.15} & \textbf{0.91}  & \textbf{55.56\%}  & \textbf{2.36} & \textbf{0.67} & \textbf{51.46\%} & \textbf{6.32} & \textbf{2.38} & \textbf{31.08\%} \\
    \bottomrule
    \end{tabular}
    }
  \label{tab:exp1}
\end{table*}

\subsection{Antigen-binding CDR-H3 Design}
\label{sec:exp2}
\noindent{\bf Data.} 
This task is to generate the specific antigen-binding CDR-H3 design, hence the antigen is a conditional input. Specifically, the CDR-H3 is only considered for generation in this task. The testing data is from~\citet{adolf2018rosettaantibodydesign}, which contains 60 complexes with different antigen types. The training data is also SAbDab dataset, but only the antigen-antibody complexes are utilized. We take the processed data by~\citet{jin2022antibody}, where the train/valid set contains $2777$ and $169$ complexes.

\begin{wraptable}{r}{0.45\textwidth}
  \centering
  \caption{AAR and RMSD evaluation results for antigen-binding antibody design task. Only CDR-H3 is for generation. The RAbD and sequence model results are taken from~\citet{jin2022antibody}, while the others are reproduced by ourselves.}
  \begin{tabular}{l cc}
    \toprule
     \textbf{Method} & \textbf{AAR$\uparrow$} & \textbf{RMSD$\downarrow$} \\
    \midrule
     RAbD & 28.6\% & - \\
     Sequence model & 32.2\% & -\\
     \midrule
     AR-GNN & 28.07\% & 2.51 \\
     RefineGNN & 34.14\% & 2.99 \\
     HERN & 33.67\% & 2.90 \\
     \midrule
     AbBERT-HMPN & \textbf{40.35\%} & \textbf{1.62}  \\
    \bottomrule
    \end{tabular}
  \label{tab:exp2}
  \vspace{-0.3cm}
\end{wraptable}

\begin{figure}[!htbp]
\centering
\includegraphics[width=0.9\linewidth]{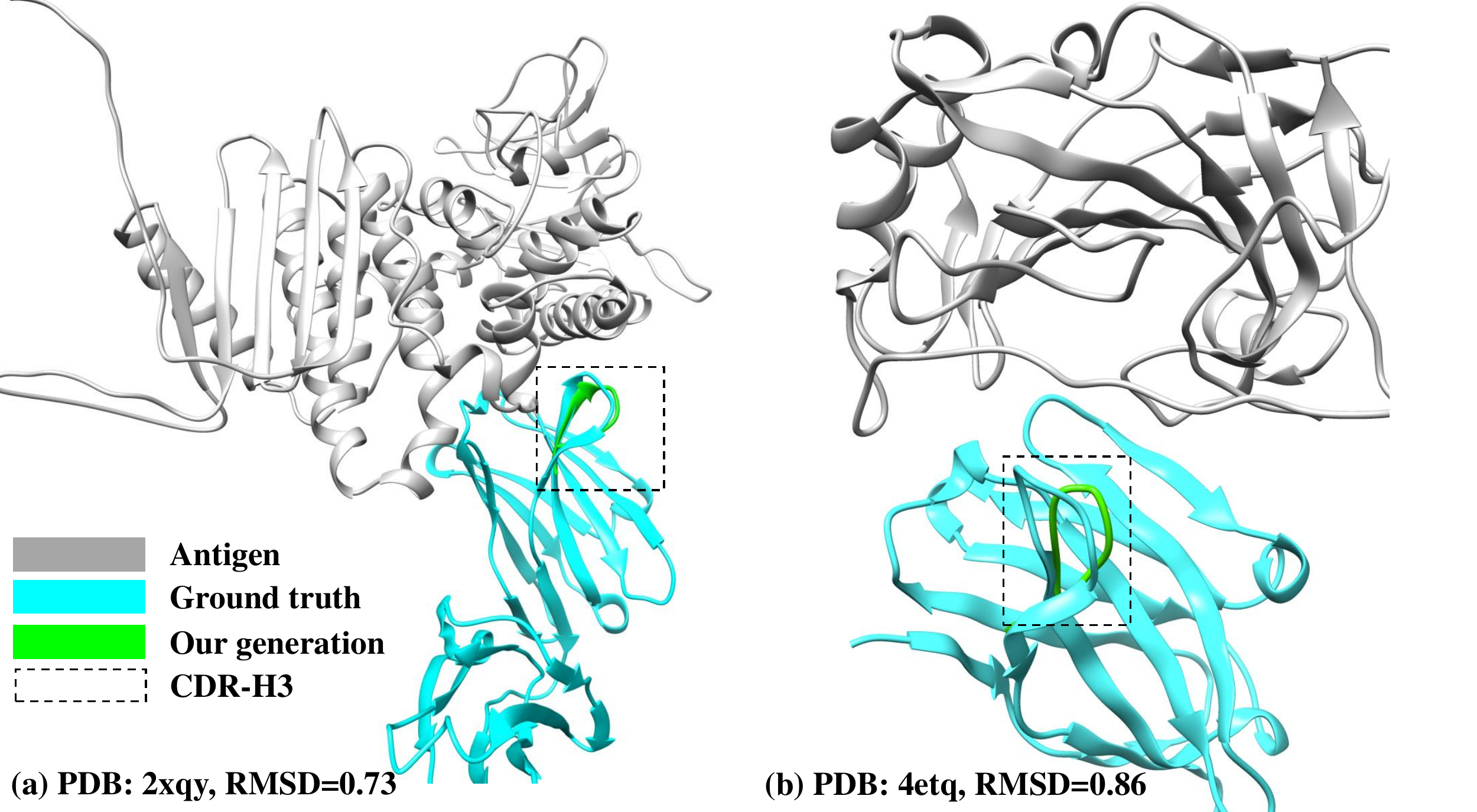}
\caption{Comparison of the ground truth and generated CDR-H3. }
\label{fig:caseStudy}
\end{figure}

\noindent{\bf Results.}
Following previous works, we use AAR for evaluation. Besides, we also take RMSD to evaluate the structure prediction accuracy. During inference, we follow~\citet{jin2022antibody} to generate 10000 CDR-H3 candidate sequences for each antibody and select the top 100 with the lowest PPL to calculate the average AAR and RMSD. The results are reported in Table~\ref{tab:exp2}. We can observe that our AbBERT-HMPN achieves the best results on AAR and RMSD with significant improvement compared to previous works, which demonstrate the strong generation ability and the potential practical value of our method. In detail, our AbBERT obtains 40.35\% AAR, which surpasses previous works by more than 6 points. Besides, the RMSD is also largely reduced by about 1.3 point to be 1.62, with almost double performance improvement. We show two cases in Figure~\ref{fig:caseStudy}, where the gray, cyan and green ribbons denote antigen, the ground-truth antibody and our generation respectively. The CDR-H3's are in the dashed box. We can observe that for these two cases, our method can generate CDRs that are close to the ground-truth with RMSD scores $0.73$ and $0.86$.


\section{Study}
In this section, we conduct different study experiments for better understanding our method. Without specific mention, the studies are all performed on antigen-binding CDR-H3 design task with $10$ epochs training.

\begin{wraptable}{r}{0.42\textwidth}
  \centering
  \caption{AAR and RMSD evaluation results for ablation study.}
  \begin{tabular}{l cc}
    \toprule
     \textbf{Setting} & \textbf{AAR$\uparrow$} & \textbf{RMSD$\downarrow$} \\
    \midrule
     AbBERT-HMPN & 1.62 \% & 40.35 \\
     \midrule
     \quad $-$AbBERT & 1.84 \% & 31.82 \\
     \quad $-$framework & 2.60 \% & 37.55 \\
     \quad $-$fusion & 1.74 \% & 36.48 \\
    \bottomrule
    \end{tabular}
  \label{tab:ablation}
  \vspace{-0.3cm}
\end{wraptable}

\subsection{Ablation Study}
We first perform ablation study to investigate the effectiveness of our proposed components.  Specifically, we conduct the following experiments: (1) AbBERT-HMPN without the pre-trained AbBERT; (2) AbBERT-HMPN without the framework encoding; (3) AbBERT-HMPN without feature fusing in the $\mathcal{H}_{seq}$, $\mathcal{H}_{seq}$. 
The results are shown in Table~\ref{tab:ablation}. From it, we have following observations: (1) removing each of the above parts will lead to performance drop (AAR becomes smaller and RMSD becomes larger), which proves the positive effect of above components. (2) Among these components, the pre-trained AbBERT contributes most to the final AAR accuracy, and the framework impacts most to the structure. Both rationally demonstrating the design of our method.

\subsection{Effect of Decoding ways}
As we mentioned the advantage of one-shot decoding, we study the autoregressive decoding and our one-shot decoding. For easy implementation, we remove the framework region is this study and then do autoregressive or one-shot training and decoding. The results are compared in Table~\ref{tab:decoding}. Besides the AAR and RMSD scores, we also provide the memory and time cost during inference to have a better comparison. We can see that our one-shot decoding not only keeps the high efficiency (memory/time cost) but also achieves better performances (AAR/RMSD).

\begin{table}[ht]
  \centering
  \caption{Performance and cost comparison between iterative and one-shot decoding.}
  \begin{tabular}{l c cc c}
    \toprule
     & \textbf{AAR$\uparrow$} & \textbf{RMSD$\downarrow$} & \textbf{Memory (G)}  & \textbf{Time (min)} \\
    \midrule
     iterative & 32.43 \% & 3.12 & 42.4 & 114 \\
     one-shot & 37.55 \% & 2.60 & 22.1 & 56 \\
    \bottomrule
    \end{tabular}
  \label{tab:decoding}
  \vspace{-0.3cm}
\end{table}

\subsection{AbBERT Incorporating Ways}
\label{sec:incor_study}
In this section, we study the different incorporating ways of the pre-trained AbBERT in our framework, or more specifically, the different tuning methods of the pre-trained AbBERT. We compare the three widely adopted methods, which are (a) fixed without tuning (serves as a feature extractor), (b) all-tuning (all parameters of the pre-trained model are finetuned), and (c) prefix-tuning (only the added prefixes are finetuned while fixing the original model). 
The results are presented in Table~\ref{tab:finetune}. From the numbers, it is obvious that our adopted prefix-tuning performs better than other two ones. This meets our expectation since the structure antibody data is in limited size and the prefix-tuning balances the best between the parameter tuning and knowledge transfer (all-tuning is easy to be overfitting and fixed version is hard to transfer for co-design). 

\begin{table}[ht]
  \centering
  \caption{Effect of different finetuning ways.}
  \begin{tabular}{l c c c}
    \toprule
     & \textbf{Fix} & \textbf{All-tuning} & \textbf{Prefix-tuning} \\
    \midrule
     \textbf{AAR$\uparrow$/RMSD$\downarrow$} & 38.62\%/1.83 & 37.82\%/1.86 & 39.68\%/1.82  \\
    \bottomrule
    \end{tabular}
  \label{tab:finetune}
\end{table}


\subsection{Effect of Pre-training}
\label{sec:pretrain_effect}

Finally, we also investigate the effect from the pre-training models. That is, we train different settings for the pre-training AbBERT model,  and evaluate the final co-design performance to see the effect. Specifically, as discussed before, we choose the mask ratio $p_m$ as a study point and set them to be $[50\%, 80\%, 100\%]$.
After finetuning for co-design, the performances are shown in Table~\ref{tab:pre-train}. 
We can see the mask ratio indeed impact the co-design performance, and the best setting is 100\% masking of the CDRs. This is also reasonable since the CDRs are about 15\% of the $V_H$ domain, the pre-training has rich context to learn a good representation for CDR prediction. Hence in the main experiment, we adopt $p_m=100\%$. 

\begin{table}[ht]
  \centering
  \caption{Effect of mask ratio $p_m$ of pre-trained AbBERT.}
  \begin{tabular}{l c cc c}
    \toprule
      & \textbf{$p_m$=50\%} & \textbf{$p_m$=80\%} & \textbf{$p_m$=100\%} \\
    \midrule
     \textbf{AAR$\uparrow$} & 36.63\% & 37.15\% & 39.68\%  \\
    \bottomrule
    \end{tabular}
  \label{tab:pre-train}
  \vspace{-0.3cm}
\end{table}

\section{Conclusions}
The antibody plays a crucial role in the therapeutic usage in out immune system. The development of the computational antibody design suffers from the limited data issue. In this work, we leverage the large-scale sequence antibody data for pre-training and then adopt the pre-trained model for co-designing the antibody sequences  and structures. With some novel designs, our method outperforms strong baselines on different tasks, such as CDR-H3 generation, sequence and structure modeling. In the future, jointly modeling the CDR-H1/H2/H3 and designing more advanced models for sequence-structure co-design are promising. 

\bibliography{reference}
\bibliographystyle{mybst}

\clearpage

\appendix
\section{Appendix}

\subsection{Pre-training Details}
\label{sec:pretrain_appendix}
The details of the OAS pre-training data are as follows. OAS currently contains over one billion sequences, from over $80$ different studies that cover diverse immune states, organisms and individuals. We firstly exclude sequences which meet criteria as mentioned in ~\cite{Bachas2022.08.16.504181}. Translated amino acid subsequences of each region are concatenated to make the full-length antibody sequences. Those sequences are clustered using MMseqs2 ~\cite{steinegger_mmseqs2_2017,steinegger_clustering_2018} with the minimum sequence identity set to 0.7 to reduce redundancy. We sample representative sequences of such clusters to make the final pre-training dataset. Tokens in framework regions are masked.

\begin{figure}[h]
    \centering 
    \includegraphics[width=0.6\linewidth]{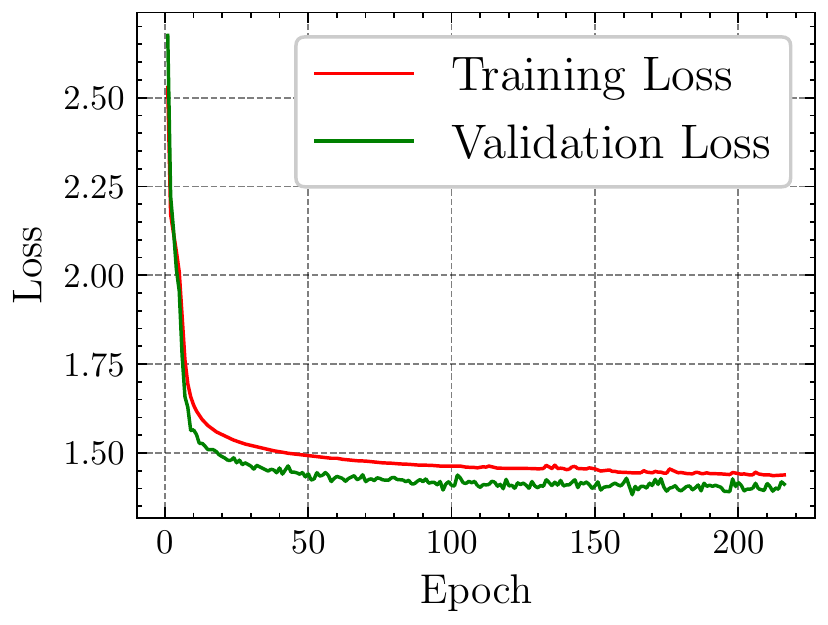}
    \caption{Pre-training loss curves on train and valid sets.}
    \label{fig:loss_curve}
\end{figure}

The AbBERT pre-training configuration is similar to BERT$_{\texttt{base}}$. We use a $12$-layer Transformer encoder with hidden dimension $768$, feed-forward network size $3072$. Adam is the optimizer with initial learning rate $3e-4$, dropout value is $0.1$. The training is conducted on 
16 V100 GPU cards. As for the performance evaluation, we plot the losses (sequence perplexity) on the training and validation sets along the training process in Figure~\ref{fig:loss_curve}. The training is stable and the loss value shows that the model has learned how to predict the correct token. 

\subsection{Model Details}
\label{sec:model_appendix}

\noindent{\bf Residue features and residue edge features.}
We follow~\citet{jin2022antibody} to build the initial residue feature and edge feature. Specifically, each amino acid consists of six features: binary polarity, binary hydrogen bond donor, binary acceptor, charge $f_c\in\{-1,0,1\}$. Another two hydropathy volume features are expanded into radial basis with interval 0.1 and 10. The total dimension is 112. As for the residue-level edge feature, a local coordinate frame $T_i = [b_i, n_i, b_i\times n_i]$ is defined, then the edge feature $f(a_i, a_j)$ between residue $r_i$ and $r_j$ is
\begin{equation}
    f = (E_p(i-j), \texttt{RBF}(||z(i,1)-z(j,1)||), T_i^T\frac{z(j,1) - z(i,1)}{||z(i,1) - z(j,1)||}, q(T_i^T T_j)),
\end{equation}
where \texttt{RBF} encodes distance, third term encodes the direction information and the last item encodes orientation information. 

\noindent{\bf Message passing network operation (MPN).}
The message passing network is easy, which is computed as follows:
\begin{equation}
    h^{l+1}(r_i) = h^l(r_i) + \sum_{j\in\mathcal{N}_i}\texttt{FFN}(h^{l}(r_i), h^{l}(r_j), f(r_i), f(r_i, r_j)),
\end{equation}
where $h^l(r_i)$ is $l$-th layer representation for residue $r_i$, \texttt{FFN} is feed-forward network. 

\noindent{\bf Structure force prediction.} 
The structure force between the $C_{\alpha}$ atoms are calculated by the nearest residues, e.g., epitope-paratope interface, 
\begin{align}
    F^{t}(i,j) & = g(h^t(Ab^p_i), h^t(Ab^p_j))\cdot(z^{t-1}(Ab^p_{i,1}) - z^{t-1}(Ab^p_{j,1})), \nonumber \\
    F^{t}(i,k) &= g(h^t(Ab^p_i), h^t(Ag^e_k))\cdot(z^{t-1}(Ab^p_{i,1}) - z^{t-1}(Ag^e_{k,1})),
\end{align}
where $g$ is feed-forward network, $t$ is the refinement step. Then the $C_{\alpha}$ coordinate is updated by the force as 
\begin{equation}
    z^t(Ab^p(i,1)) = z^{t-1}(Ab^p(i,1)) + \frac{1}{n}\sum_{j\neq i}F^t(i,j) + \frac{1}{m}\sum_kF^t(i,k).
\end{equation}

For other atoms, the force is calculated among the same residue. For example, for atom $Ab^p_{i,j}$ and $Ab^p_{i,k}$, the force calculation and $Ab^p_{i,j}$ coordinate update is
\begin{align}
    F^{t}(ij,ik) &= g(h^t(Ab^p_{i,j}), h^t(Ab^p_{i,k}))\cdot(z^{t-1}(Ab^p_{i,j}) - z^{t-1}(Ab^p_{i,k})), \nonumber \\
    z^t(Ab^p(i,j)) &= z^{t-1}(Ab^p(i,j)) + \frac{1}{n_i}\sum_kF^t(ij,ik).
\end{align}

\noindent{\bf Training details.} 
For training the co-design model, we use Adam optimizer with seldom hyperparameter search. The learning rate is among $[1e-4, 5e-4]$, epoch ranges in $[10,20]$. We set number of prefix tokens to be $5$, dropout value $0.1$.

\end{document}